# Steroid Receptors and Vertebrate Evolution

Michael E. Baker
Division of Nephrology-Hypertension
Department of Medicine, 0693
University of California, San Diego
9500 Gilman Drive
La Jolla, CA 92093-0693

Correspondence to:
M. E. Baker. E-mail: mbaker@ucsd.edu

**Abstract** Considering that life on earth evolved about 3.7 billion years ago, vertebrates are young, appearing in the fossil record during the Cambrian explosion about 542 to 515 million years ago. Results from sequence analyses of genomes from bacteria, yeast, plants, invertebrates and vertebrates indicate that receptors for adrenal steroids (aldosterone, cortisol), and sex steroids (estrogen, progesterone, testosterone) also are young, with an estrogen receptor and a 3-ketosteroid receptor first appearing in basal chordates (cephalochordates: amphioxus), which are close ancestors of vertebrates. Through gene duplication and divergence of the 3-ketosteroid receptor, receptors that respond to androgens, glucocorticoids, mineralocorticoids and progestins evolved in vertebrates. Thus, an ancestral progesterone receptor and an ancestral corticoid receptor, the common ancestor of the glucocorticoid and mineralocorticoid receptors, evolved in jawless vertebrates (cyclostomes: lampreys, hagfish). This was followed by evolution of an androgen receptor, distinct glucocorticoid and mineralocorticoid receptors and estrogen receptor-α and -β in cartilaginous fishes (gnathostomes: sharks). Further evolution of mineralocorticoid signaling occurred with the evolution of aldosterone synthase in lungfish, a forerunner of terrestrial vertebrates. Adrenal and sex steroid receptors are not found in echinoderms: and hemichordates, which are ancestors in the lineage of cephalochordates and vertebrates. The evolution of steroid receptors at key nodes in the evolution of vertebrates argues for an important role for steroid receptors in the evolutionary success of vertebrates, considering that the human genome contains about 22,000 genes, which is not much larger than genomes of invertebrates, such as *Caenorhabditis elegans* (~18,000 genes) and *Drosophila* (~14,000 genes).

**Keywords:** Steroids, Vertebrate Evolution, Amphioxus, Ancestral Estrogen, Cambrian Explosion, Snowball Earth

## 1. Vertebrates evolved late in the evolution of life on earth

Life in the form of cynanobacteria is thought to evolved on earth about 3.7 billion years ago, with single cell eukaryotes, fungi (e.g. yeast, ancestors of modern *Saccharomyces cerevisiae*) evolving about 2 billion years ago [1]. Molecular clock analyses suggest that multicellular animals (metazoans) originated from 800 to 1,000 million years ago [2-4]. Sponges (Porifera) [5] and *Trichoplax adhaerens* (Placazoa) [6, 7], one of the simplest multicellular animals, are modern representatives of basal metazoans [8]. More complex body plans are found in cnidarians [9], which are diploblasts, containing two germ layers, an ectoderm and endoderm. Modern examples are jellyfish, corals, sea anemones and comb jellies. The evolution of Bilateria: bilaterally symmetrical animals that are triploblasts, animals with three germ layers: a mesoderm as well as ectoderm and endoderm, followed. Invertebrates are diverse, containing insects, such as *Drosophila*, the nematode *Caenorhabditis elegans*, and mollusks, such as scallops, clams, snails and octopus. Vertebrates evolved in the deuterostome line, a sister clade of protostomes, which contains invertebrates [10, 11]. Deuterostomes are a much smaller clade of animals. They include echinoderms (sea urchins) and chordates (sea squirts, lancelets, and vertebrates). Thus, vertebrates and other chordates are young in comparison to the evolution of life and to the evolution of multicellular animals. Vertebrate fossils first appeared in the Cambrian from 542 to 515 million years ago [4, 11]. However, microscopic animal ancestors of vertebrates evolved from the 250 to 400 million years prior to the Cambrian explosion [2, 3], when periods of extreme climate called "Snowball Earth" stimulated the evolution of many transcription factors, including steroid receptors [12, 13]. The basis for the explosive increase in the diversity of large multicellular animals in the Cambrian is still unresolved [4, 13-15].

## 2. Nuclear Receptors for steroids and other lipophilic ligands evolved in multicellular animals

Nuclear receptors are a large and diverse family of transcription factors that contain receptors for adrenal steroids (glucocorticoids, mineralocorticoids) and sex steroids (estradiol, progesterone, testosterone) (Figure 1), as well as other lipophilic hormones including thyroid hormone, retinoic acid, 1,25-dihydroxyvitamin D3, oxysterols, bile acids and ecdysone [5, 16-19]. Adrenal and sex steroid receptors are a separate clade distinct from other nuclear receptors [20, 21] including the ecdysone receptor, which is an important insect steroid receptor. Some nuclear receptors, which do not have a bona fide biological ligand, are called orphan receptors [22-24]. Ligands for some of these receptors may not have been discovered, while some receptors may not require a ligand for biological activity.

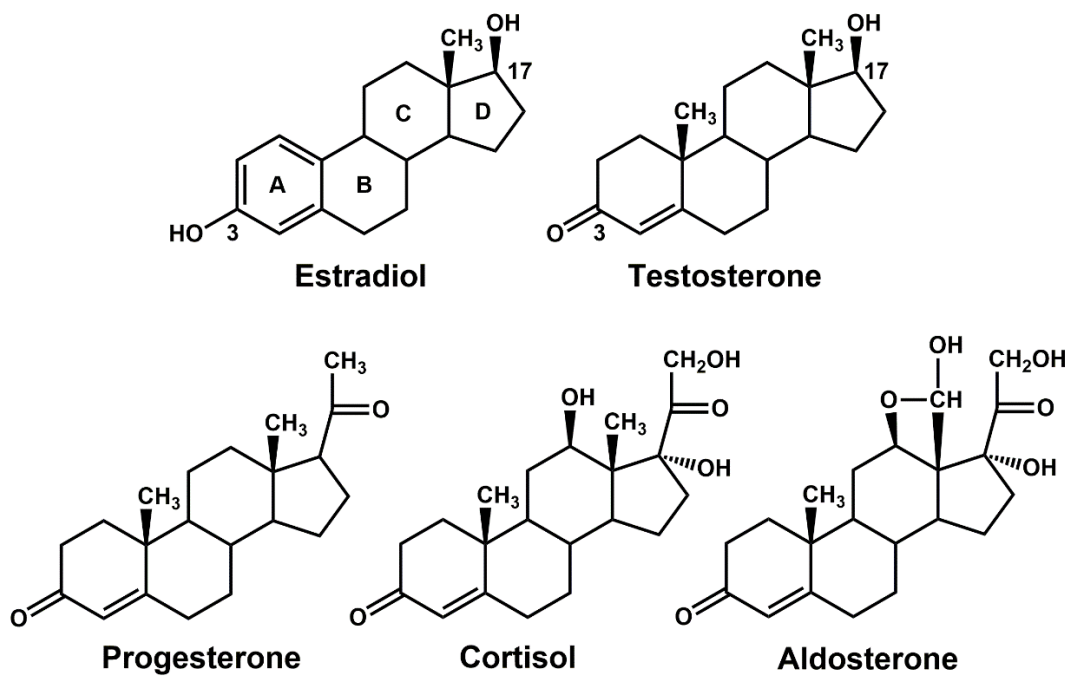

**Figure 1** Structures of vertebrate adrenal and sex steroids.
Estradiol has an aromatic A ring with a C3 hydroxyl and lacks a C19 methyl group. Testosterone, progesterone, cortisol and aldosterone contain an A ring with an unsaturated bond between C4 and C5 and a C3-ketone, and a C19 methyl group. Estradiol is important in female and male physiology [50, 107, 108]. Aldosterone is the main physiological mineralocorticoid in vertebrates [42, 67, 75]. Cortisol is the main physiological glucocorticoid in humans [42]. Testosterone and progesterone are reproductive hormones in males and females, respectively. Progesterone is an antagonist for the MR in humans [109], alligators and *Xenopus* [81] and an agonist for fish MRs [81-83, 95] and chicken MR [81].

Sequence analysis indicates that nuclear receptors are not present in yeast or plants. Nuclear receptors have been found in basal metazoans: sponges (Porifera) [5] and *Trichoplax adhaerens* (Placazoa) [6, 25-27], which contain two and four nuclear receptors, respectively. Trichoplax has an estrogen related receptor γ (ERRγ), a retinoid X receptor (RXR), a chicken ovalbumin upstream promoter receptor (COUP) and an Hepatocyte Nuclear Factor 4 receptor (HNF4). ERR and COUP are orphan receptors. During the evolution of multicellular animals, nuclear receptors duplicated, diverged and their number expanded considerably [5, 28]. Interestingly, there are large differences in the number of nuclear receptors in various invertebrates and vertebrates. For example, there are 17 nuclear receptors in the cnidarian starlet sea anemone *Nematostella vectensis* [9], 270 nuclear receptors in the nematode *C. elegans* [29], 18 nuclear receptors in *Drosophila* [30], and 48 nuclear receptors in humans [31]. Thus, from simple origins, nuclear receptors evolved into evolved into a diverse family of transcription factors that are essential to multicellular animals [28].

Interestingly, despite the absence of steroid receptors in yeast, transfection of the yeast *Saccharomyces cerevisiae* with either human estrogen receptor α (ERα) [32] or rat glucocorticoid receptor (GR) [33] yields cells that can respond to estradiol and dexamethasone, respectively. Moreover, transfection of the plant Arabidopsis with rat GR [34] leads to cells that are activated by dexamethasone. This indicates that the proteins needed for transcriptional activation of ERα and the GR evolved before multicellular animals.

Although nuclear receptors evolved in basal metazoans, as discussed below, adrenal and sex steroid receptors evolved much later at important nodes, preceding and during the evolution of vertebrates, beginning with the first steroid receptors in cephalochordates [35-37], which are marine invertebrates that are considered to be close relatives of the vertebrates [38-40]. Cephalochordates contain a notochord [39, 40], which extends the length of the body, and is an important evolutionary innovation that becomes part of the vertebral column in vertebrates. Further evolution of steroid receptors to respond to steroids with diverse structures occurred in the first jawless vertebrates [41] and jawed cartilaginous fishes [17, 42].

## 3. Adrenal and sex steroid receptors evolved in chordates

In the 1990s as more eukaryote sequences became available, the origins of nuclear receptors including adrenal and sex steroid receptors began to be investigated [19, 20, 43]. Phylogenetic analysis, based on available steroid receptor sequences in vertebrates, predicted that the ER was the ancestral vertebrate steroid receptor, with an origin in a close vertebrate ancestor, such as amphioxus [20, 44, 45].

## 4. An estrogen receptor and 3-ketosteroid receptor evolved in amphioxus.

The cloning of an ortholog of the vertebrate ER from amphioxus was an important advance in deciphering the evolution of steroid receptors [35-37] (Figure 2). Unexpectedly, amphioxus ER does not bind estradiol or other steroids, such as progesterone, cortisol, corticosterone, aldosterone and testosterone. Amphioxus also contains a 3-ketosteroid receptor (SR) that is activated by estradiol and estrone, suggesting that amphioxus SR evolved from a duplication of amphioxus ER [35, 36], with amphioxus ER losing estrogen-binding activity. This activity, however, was retained in amphioxus SR, which is not activated by progesterone, cortisol, corticosterone, aldosterone or testosterone.

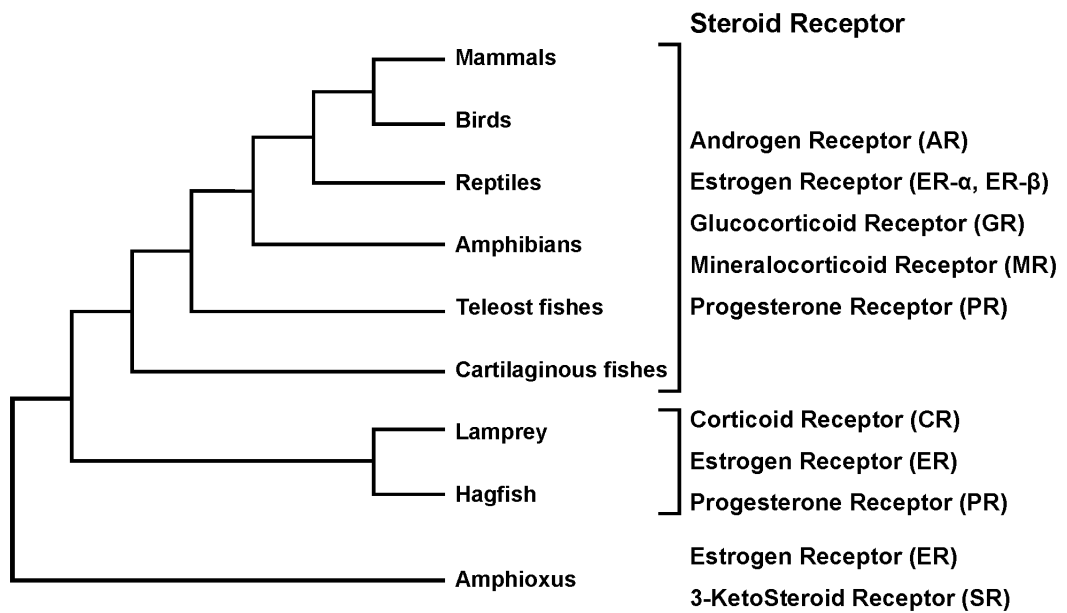

**Figure 2 Steroid receptors in amphioxus and vertebrates**
The ER and SR evolved in amphioxus. The CR and PR evolved in lampreys and hagfish, which are jawless fishes (cyclostomes) at the base of the vertebrate line. The AR, a separate MR and GR and ERα and ERβ first appear in cartilaginous fishes, which are forerunners of bony fishes and lobe-finned fish (coelacanths and lungfish). Lobe-finned fish are forerunners of tetrapods (land vertebrates).

**5. What was the ancestral estrogen?**

Different molecule(s) have been proposed to be the ancestral estrogen [35, 41, 46-48]. Amphioxus contains aromatase (CYP19) an enzyme necessary for synthesis of the aromatic A ring of estradiol [49], supporting the hypothesis that estradiol or estrone or both were ancestral estrogens [35, 41]. Indeed, the presence of aromatase in amphioxus, but not in invertebrates [17] supports the hypothesis invertebrates do not contain an estrogen-activated ER ortholog.

However, an intriguing property of vertebrate ERs is their promiscuous response to chemicals, which suggests that chemicals with diverse structures that lack an aromatic A ring may have been ligands for amphioxus ER and SR [50, 51]. For example, in addition to estradiol, human ER is activated by nM concentrations of Δ5-androstenediol and 5α-androstanediol [50, 52, 53] (Figure 3). Both steroids are attractive alternative physiological estrogens because their synthesis is simpler than that of estradiol [17, 54] as neither Δ5-androstenediol nor 5α-androstanediol has an aromatic A ring (Figure 3). Indeed, Δ5-androstenediol is synthesized in one step from dehydroepiandrosterone (DHEA) (Figure 4). Δ5-androstenediol is a physiological estrogen in mammals with important actions in mammalian brain [55-57] and other organs [58, 59]. This supports a role for Δ5-androstenediol in ancestral chordates, such as amphioxus [38]. Another possible ancestral estrogen is 27-hydroxycholesterol which activates human ER [60-62] (Figure 3). Indeed, 27-hydroxycholesterol is the physiological estrogen in humans with the simplest synthetic pathway [50, 51] (Figure 4).

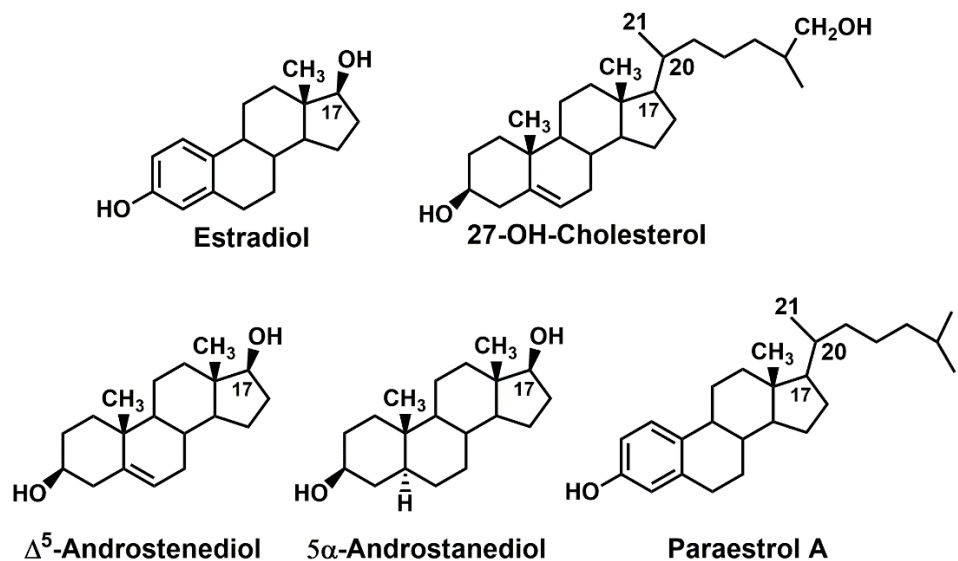

**Figure 3 Potential ancestral estrogens.** Estradiol contains an aromatic A ring with a C3-hydroxyl group. In contrast, Δ5-androstenediol, 5α-androstanediol and 27-hydroxycholesterol contain a cyclohexane A ring, with a 3β-hydroxyl and a C19 methyl group. Δ5-androstenediol and 27-hydroxycholesterol contain a B ring with an unsaturated bond between C5 and C6. Δ5-androstenediol, 5α-androstanediol and 27-hydroxycholesterol are physiological estrogens [50-52, 61, 62, 107]. Paraestrol A, which has an aromatic A ring, is synthesized by corals binds to the ancestral SR [65, 66].

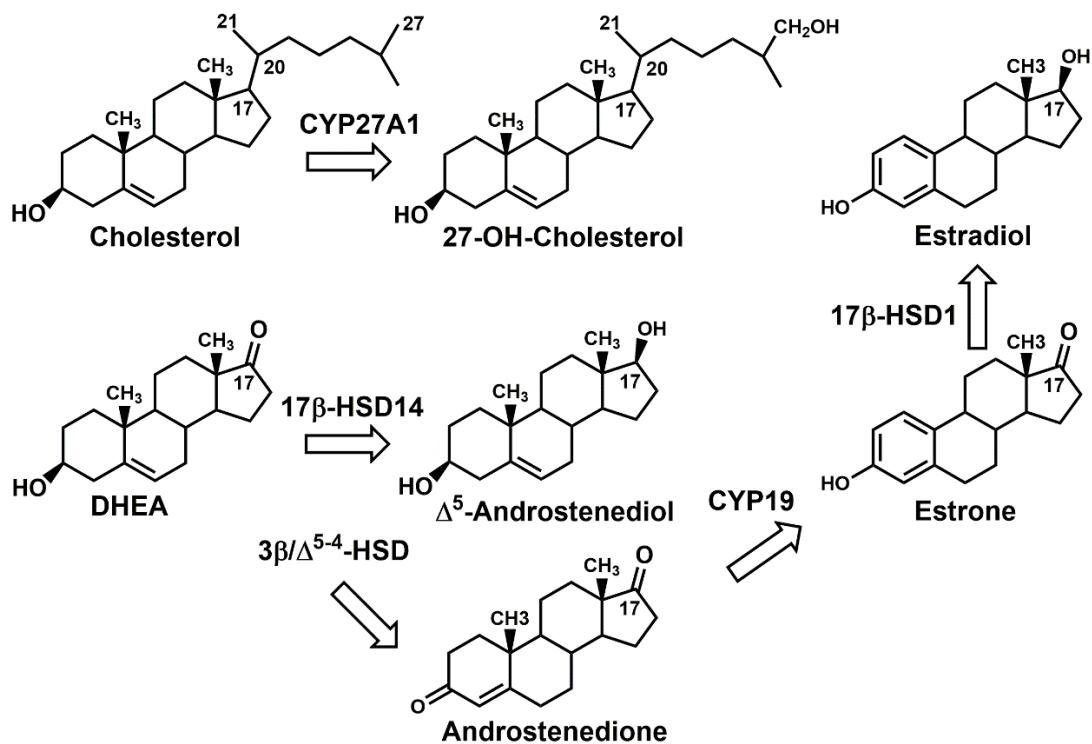

**Figure 4 Comparison of estradiol, Δ5-androstenediol and cholesterol synthesis.** Estradiol is synthesized in three steps (3β/Δ$^{5-4}$-hydroxysteroid dehydrogenase (3β/Δ$^{5-4}$-HSD), cytochromeP450 19 (CYP19), 17β-hydroxysteroid dehydrogenase type 1 (17β-HSD1)) from dehydroepiandrosterone (DHEA) [17, 38, 54]. In contrast, Δ5-androstenediol is synthesized from dehydroepiandrosterone (DHEA) in one step (17β-hydroxysteroid dehydrogenase type14 (17β-HSD14)) [38, 107]. 27-hydroxycholesterol is synthesized in one step from cholesterol [54, 107].

Interestingly, animals such as sponges [63] and coral [64] synthesize steroids with an aromatized A ring and a C17 side chain and could have activated the ancestral ER [65]. Markov et al synthesized one of these steroids, paraestrol A (Figure 3), which has been found in coral [27, 64, 65], and reported that paraestrol A is a transcriptional activator of an ancestral SR [65]. The synthesis by coral of a steroid with an aromatic A ring that was not catalyzed by vertebrate CYP19 raises the possibility that invertebrates could synthesize other aromatized steroids that were transcriptional activators of an ancestral ER [65, 66].

## 6. Diversification of 3-ketosteroid receptors in jawless fishes: evolution of progesterone and corticoid receptors

An important advance in understanding the timing of the evolution of adrenal and sex steroid receptors in vertebrates was the sequencing by Thornton [41] of a PR, a CR, which is an ancestor of the MR and GR, as well as an ER, in lamprey, which is a basal vertebrate with more complicated development and life history than amphioxus. The evolution of the PR and CR in a jawless vertebrate through duplication of an ancestral SR [17, 41, 67] added 3-ketosteroids to the repertoire of physiological ligands, providing more precise regulation of complex physiological pathways, including reproduction, immune responses, electrolyte homeostasis and stress responses, in jawless vertebrates and their descendants in cartilaginous fishes, ray-finned fishes and terrestrial vertebrates.

The physiology of steroids in lamprey is only beginning to be understood. Close et al. found that 11-deoxycortisol functioned in lamprey as both a mineralocorticoid and glucocorticoid through activation of the CR [68]. The physiological actions of progestins in lamprey are still being elucidated [69]. Interestingly, both progesterone and 15α-hydroxyprogesterone circulate in lamprey serum [69, 70]. Although an androgen receptor has not been found in lamprey, both testosterone and 15α-hydroxytestosterone are synthesized by lampreys [70, 71]. The role of 15α-hydroxysteroids in lamprey is novel and needs further study.

## 7. Evolution of orthologs of androgen, glucocorticoid and mineralocorticoid receptors and estrogen receptors α and β in cartilaginous fishes.

Further diversification of the steroid receptor family occurred in cartilaginous fishes (Gnathostomes: sharks), in which a distinct AR ortholog evolved from a duplicated PR, and distinct MR and GR orthologs evolved from a duplicated CR [17, 67]. The evolution in cartilaginous fishes of orthologs of the AR, GR and MR completed the evolution of the 3-ketosteroid receptors.

Further diversification of estrogen signaling also occurred in cartilaginous fishes in which ERα and ERβ first evolved [17, 72, 73]. In mammals, estradiol activation of ERα and ERβ increases the versatility of estrogen-mediated responses in vertebrates. In some tissues, activation of ERα by estradiol is antagonistic to activation of ERβ by estradiol {Paech, 1997 #153;Weihua, 2000 #154}. that is, activation of ERα promotes cell growth, while activation of ERβ counteracts cell proliferative by ERα, a mechanism that has been characterized as a Yin-Yang model for ERα and ERβ {Lindberg, 2003 #156}.

Together, duplications of CR, PR and ER in an ancestral cartilaginous fish completed the evolution of the five classes of adrenal and sex steroid receptors (Figure 2) and increased the versatility of steroid-activated responses in vertebrates.

However, as discussed next, this was not the end of the evolution of steroid hormone signaling in vertebrates, which continued with the evolution of CYP11B2, aldosterone synthetase, a key enzyme in the evolution of terrestrial vertebrates (Figure 5) [17, 54, 74-76].

**8. Aldosterone synthesis and the conquest of land**.

Aldosterone, the physiological mineralocorticoid in humans and other terrestrial vertebrates was important in evolution of vertebrates from an aqueous environment to an aerobic environment on land. Aldosterone activation of the MR in the kidney maintains salt and water homeostasis, which is critical for controlling blood pressure, as well as to prevent dehydration in animals living out of water [75].

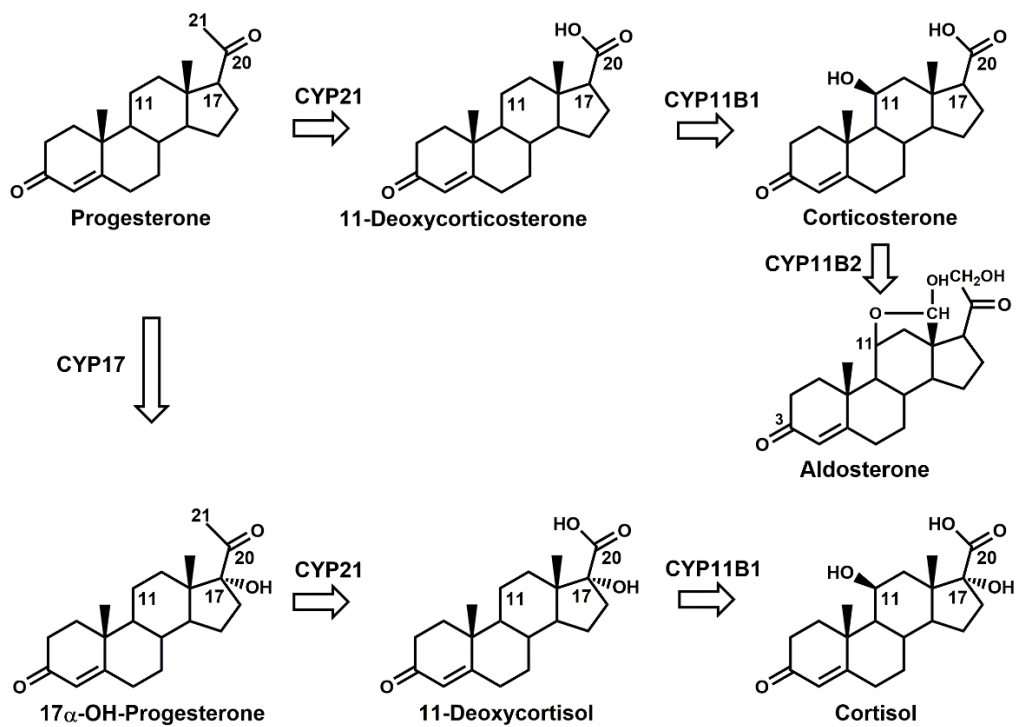

**Figure 5. Pathways for the synthesis of aldosterone and cortisol.**
Progesterone is hydroxylated at C21 to form 11-deoxycorticosterone, which is hydroxylated at C11 to form corticosterone. Corticosterone is hydroxylated cat C18 followed by oxidation of the C18 hydroxyl to form aldosterone. In a second pathway, progesterone is hydroxylated at C17 and then hydroxylated at C21 to form 11-deoxycortisol, which is hydroxylated at C11 to form cortisol [17, 42, 75].

Although aldosterone is a strong activator of the MR in lamprey and hagfish, aldosterone is not found in either lamprey or hagfish serum [77]. Moreover, although aldosterone is a strong activator of cartilaginous fish MR [77, 78] and ray-finned fish MRs [79-84], aldosterone has not been found in either cartilaginous fishes or ray-finned fishes [67, 85]. Aldosterone synthesis requires CYP11B2, which has not been found in fish [85]. Aldosterone first appears in lungfish, forerunners of terrestrial vertebrates [67, 76, 86], well after in the evolution of corticosteroids in lampreys, hagfish and cartilaginous fishes [17, 75].

An important role for aldosterone and the MR in the conquest of land by vertebrates came from the work of Homer Smith, a comparative physiologist who studied lungfish during his travels to Africa [75, 87]. Lungfish can live in aqueous and aerobic environments, leading Smith to propose that the evolution of a mechanism to prevent dehydration in lungfish on land was important in the conquest of land by terrestrial vertebrates. The evolution of CYP11B2, leading to aldosterone synthesis in lungfish, provides a mechanism for tight control of electrolyte levels in serum. By maintaining homeostasis in this "Internal Environment", aldosterone activation of the kidney MR facilitated the conquest of land by vertebrates [75].

## 9. Perplexing questions:
### A. What is the ligand for the mineralocorticoid receptor in ray-finned fishes?

Ray-finned fish do not synthesize aldosterone, raising the question as to what is the physiological ligand for the MR in ray-finned fish? Cortisol and 11-deoxycorticosterone have been proposed as physiological mineralocorticoids (Figure 5) [82, 88-94]. However, complicating the identity of the ligand for the MR in ray-finned fishes is evidence that progesterone (Figure 1) is a transcriptional activator of several ray-finned MRs [81-83, 95]. Although progesterone is thought of as a female reproductive hormone [96], progesterone also is active in males [97]. Moreover, progesterone is a precursor of aldosterone and cortisol (Figure 5), and thus, has a much simpler synthesis. This suggests that in addition to activating the PR, progesterone also may regulate transcriptional activation of the MR in ray-finned fish [67, 81, 95].

### B. Physiological activity of ray-finned fish mineralocorticoid receptor?

Unexpectedly, cortisol has been found to regulate sodium uptake in zebrafish through transcriptional activation of the GR [98, 99]. Mineralocorticoid activity of the cortisol via activation of the GR, instead of the MR, in fish [92, 98, 99] is perplexing. Indeed, the physiological function of ray-finned fish MR is incompletely understood. Thus far, ray-finned fish MR appears to have physiological function(s) in fish brain [92, 93, 100]. Interestingly, mammalian MR is expressed in the brain, where MR function is still being elucidated [67, 101-103]. Investigation of fish MR function may provide important insights into the function of mammalian MR in the brain.

**C. Function of progesterone activation of the chicken mineralocorticoid receptor?**

Chicken MR is activated by progesterone [81]. Thus far, chicken MR is the only terrestrial vertebrate MR that is activated by progesterone [67]. The function of this novel activity of progesterone in chickens remains to be elucidated.

**10. Steroid receptors were important in vertebrate success and in survival during environmental cataclysms**.

The human genome contains about 22,000 genes [31] which is not much more than the ~18,000 genes in *C. elegans* [29] and ~14,000 genes in *Drosophila* [104]. The small increase in the number of human genes compared to *C. elegans* and *Drosophila* is perplexing considering the vast differences in differentiation, development and reproduction between humans and *C. elegans* and *Drosophila*. The evolution of different classes of adrenal and sex steroid receptors beginning with the ER and SR in amphioxus, followed by the CR and PR in jawless fish and AR, MR and GR cartilaginous fish supports the hypothesis that steroid receptors were important to the success of vertebrates at key nodes in the diversification of vertebrates by providing more flexible regulation of diverse physiological pathways during differentiation, development, reproduction, immune responses, electrolyte homeostasis and the stress response.

The complexity of vertebrate physiology also makes vertebrates sensitive to major cataclysms that disrupt the environment. The flexibility provided by adrenal and sex steroid receptors in regulating vertebrate physiology may have been important in vertebrate survival of major environmental cataclysms, such as the mass extinction at the end of the Permian about 252 million years ago, which saw the loss of up to 96% of all marine species and 70% of terrestrial species [105, 106].

**Acknowledgement.**

Supported by Research Fund #3096.